\documentclass[%
 reprint,
 amsmath,amssymb,
 aps,
]{revtex4-2}

\usepackage{graphicx}
\usepackage{dcolumn}
\usepackage{bm}

\usepackage{braket}
\usepackage{mathcomp}
\usepackage{gensymb}
\usepackage{here}
\usepackage{amsmath}
\usepackage{lineno}

\usepackage{ulem}
\usepackage{xcolor}
\DeclareRobustCommand{\erase}{\bgroup\markoverwith{\textcolor{red}{\rule[.5ex]{2pt}{0.4pt}}}\ULon}

\begin{document}



\title{Post-selected enhancement of phase sensitivity in a two-path interferometer}

\author{Ryuya Fukuda}
\author{Katsuki Tanaka}%
\author{Masataka Iinuma}%
\author{Holger F. Hofmann}%


\affiliation{%
 Graduate School of Advanced Science and Engineering, Hiroshima University, Kagamiyama 1-3-1, Higashi Hiroshima, 739-8530, Japan
}%






\begin{abstract}
Post-selected metrology can exceed the standard quantum limit of single photon interference by concentrating all of the available Fisher information in a subset of post-selected outcomes. Here we apply this principle to a standard two-path interferometer, activating the Fisher information concentrated in the low probability output port by transferring the phase sensitivity to the circular polarizations of the photon. We obtain a phase gradient of the post-selected interference fringe that corresponds to that of a four-photon NOON state, providing a direct comparison of the performance of multi-photon interference with single-photon post-selection. 


\end{abstract}

\maketitle


Quantum metrology makes use of the non-classical properties of measurement statistics to enhance the sensitivity of parameter estimation \cite{Bra94,Fuj95,Gio04,Gio06,Gio11,Hof12,Pez18}. On the applied side, the workhorse of quantum metrology is the two-path interferometer, where the parameter to be estimated is the phase shift between the two arms \cite{ Dow08,Hof09,Aba11,Abb16,Pol20}. Up to now, the focus has been on the use of non-classical input states, where single photon interference defines the standard quantum limit and enhancements are achieved by multi-photon interferences, ideally represented by N-photon NOON states, where the input state of light is in an equal superposition of all photons in one path or in the other \cite{Dow08,Mit04,Wal04,Afe10,Slu17}. On the formal side, the resource of phase estimation can be quantified by the quantum Fisher information. For pure states, this resource is determined by the uncertainty of the generator of the phase shift. In the case of a two-path interferometer, this is the uncertainty of the photon number difference between the paths, which confirms that no states can achieve a greater phase sensitivity than the NOON states \cite{Bra94,Hof09}. 

The general theory seems to permit no further improvements. If the physical resource is identified with the number of input photons, the quantum Fisher information is determined by the quantum statistics of the input state and NOON states achieve the highest possible Fisher information per photon \cite{Com14,Fer14}. However, one might argue that the number of input photons is not the relevant resource in a phase estimation procedure. The amount of data that needs to be recorded and stored depends on the number of photons that are actually detected, not on the (often unknown) number of input photons \cite{Har17}. It is therefore useful to consider the application of post-selection to quantum metrology  \cite{Hof12,Sta10,Bru10,Xu13,Yam13,Jor14,Pan15,Sin17,Zhu25}.
As was shown theoretically, post-selection can concentrate the Fisher information of the input state in a small fraction of the output state \cite{Arv20,Yan22,Arv24}. In such scenarios, it is not necessary to keep track of the post-selection probability, since the Fisher information is observed in the conditional probabilities of the post-selected outcomes. For single photons, this means that more than one output port must be post-selected, so that the Fisher information can be observed in the difference between the count rates of the remaining output ports. The  quantum Fisher information per detected photon can then be much higher than the Fisher information per input photon, as demonstrated experimentally for the rotation angles of photon polarization \cite{Lup22}. The implications should be clear - post-selection can enhance single photon phase sensitivity beyond the standard quantum limit by selectively detecting the photons that carry the highest concentration of quantum Fisher information. But is it possible to apply the same principle to a conventional two-path interferometer? If only the photons in one output port are detected, the conditional probability of the outcome is $1$ and the phase sensitivity is only recovered by treating non-detection events as outcomes, indirectly detecting the photons in the other port by their absence. The post-selection process used in \cite{Lup22} solved this problem by a partial attenuation that changed the balance between two components of the photon state. Even though that work demonstrated impressively that the negativity of a specific Kirkwood-Dirac quasiprobability enhances quantum metrology, it is difficult to apply partial attenuation to path superpositions. Here, we show that the problem can be solved by transferring the phase information to the circular polarization of the photons, allowing us to observe interference fringes in only one of the two output ports of the two-path interferometer without the use of partial attenuation. The phase sensitivity of the post-selected photons can be observed without any reference to the total input photon number, so that the result is robust against photon losses or intensity fluctuations in the input \cite{Slu17}.

In a recent result, we experimentally observed anomalous path uncertainties in single photon interference, demonstrating that the post-selection of the low probability outputs is correlated with a quantitative measure of squared photon number difference between the paths that can be much larger than one \cite{Fuk26}. As we noted in that work, this anomalous value of photon number difference between the paths could be used to enhance quantum metrology since it is characterized by the same negative Kirkwood-Dirac terms that were found to enhance phase sensitivity in \cite{Lup22}. However, the implementation is not straightforward. In order to concentrate the Fisher information in only one output port of the interferometer, it is necessary to transfer the phase information to an ancillary degree of freedom. We achieve this task by applying an opposite phase bias to the right and left circular polarizations of the photons. Near destructive interference, this phase bias ensures that the logarithmic derivatives of the two polarizations have opposite sign. Post-selected metrology then achieves a sensitivity determined by the squared value of these logarithmic derivatives, corresponding to the squared imaginary weak values of the photon number differences between the paths. In our experiment, we achieve a phase gradient of circular polarization that corresponds to a gradient that would require a four photon NOON state when post-selection is not used. Since we have mapped the phase sensitivity onto an ancillary system, we retain the characteristics of an interference fringe obtained from two symmetric outputs, allowing for a straightforward comparison of the result with multi-photon fringes. As we discuss in detail in the following, the ancillary use of photon polarization activates the anomalous conditional uncertainties of photon number difference between the paths, effectively replacing the input state uncertainty with the anomalous conditional uncertainty as the ultimate limit of phase sensitivity \cite{Ji26}.

Let us start with a theoretical discussion of the problem. We consider a conventional two-path interferometer. The quantum state of a single photon in path $1$ is given by $\ket{1}$ and the state of the photon in path $2$ is given by $\ket{2}$. The photon number difference between the two paths is given by an operator
\begin{equation}
    \hat{A}=\ket{1}\bra{1}-\ket{2}\bra{2}
\end{equation}
and a phase shift between the two paths is given by the unitary transformation
\begin{equation}
    \hat{U}(\phi) = \exp(-i\frac{1}{2} \hat{A} \phi).
\end{equation}
With this terminology, the quantum Fisher information of single photon interference is given by the uncertainty of $\hat{A}$, which has a maximal value of one for an equal superposition of both paths. As shown in \cite{Arv24}, post-selection of a set of outcomes $\hat{E}$ results in a quantum Fisher information of 
\begin{equation}
\label{eq:EFI}
    I_{ps} = \frac{\bra{\psi}\hat{A} \hat{E} \hat{A} \ket{\psi}}{\bra{\psi}\hat{E}\ket{\psi}} - \left|\frac{\bra{\psi} \hat{E} \hat{A} \ket{\psi}}{\bra{\psi}\hat{E}\ket{\psi}} \right|^2.
\end{equation}
Here, the state $\ket{\psi}$ is defined at a phase shift of $\phi=0$. Since the effects of post-selection depend on the relation between the operator $\hat{E}$ representing the post-selected outcomes and the initial state $\ket{\psi}$, the post-selected quantum Fisher information is phase dependent and the state $\ket{\psi}$ should be chosen with an optimal operating point at $\phi=0$ in mind.

Post-selection can enhance single photon phase sensitivity beyond the standard quantum limit of $\Delta A^2=1$ when the first term in Eq. (\ref{eq:EFI}) is larger than one. Since the eigenvalues of $\hat{A}$ are limited to $\pm 1$, this requires negative quasiprobabilities for these eigenvalues \cite{Arv20,Arv24}. 
Negative quasiprobabilities are therefore necessary for any enhancement of post-selected phase sensitivity beyond the standard quantum limit, as demonstrated in \cite{Lup22}. However, they are not sufficient to achieve such an enhancement, since the subtraction of the second term can undo the enhancement described by the first term. The reason for this subtraction becomes obvious when considering the post-selection of pure states. In that case, the post-selected Fisher information must be zero, because the output state has no phase dependence left. This is why it is so difficult to apply post-selection to a two-path interferometer - even though the selection of a single output port is sufficient to obtain a negative quasiprobability, the first term in Eq.(\ref{eq:EFI}) will be canceled by the subtraction of the absolute value squared of the weak value of $\hat{A}$.  

For a balanced two-path interferometer, the output ports are described by equal superpositions of the paths, $\ket{\pm}=(\ket{1}\pm \ket{2})/\sqrt{2}$. Since the action of $\hat{A}$ on these states flips the sign of these states, a post-selection of $\hat{E}=\ket{-}\bra{-}$ is characterized by a  quantum Fisher information of 
\begin{equation}
\label{eq:pure}
    I_{ps} = \frac{\braket{\psi|+} \braket{+|\psi}}{\braket{\psi|-}\braket{-|\psi}} - \left| \frac{\braket{\psi|-} \braket{+|\psi}}{\braket{\psi|-}\braket{-|\psi}} \right|^2.
\end{equation}
If the input state $\ket{\psi}$ is defined in the two dimensional Hilbert space of the path states, the post-selected Fisher information would indeed be zero because the inner products $\braket{\psi|-}$ in the weak values of $\hat{A}$ would cancel. However, such a cancellation does not apply if $\ket{\psi}$ is defined in a product space of path and polarization, so that $\braket{-|\psi}$ is a polarization state conditioned by the projection on $\bra{-}$. This analysis provides us with the key insight that makes our experiment possible. We can activate the post-selected phase sensitivity in the output port $\ket{-}$ of our interferometer by introducing only a slight entanglement between polarization and paths, so that most of the input state is still in the state $\ket{+}$, orthogonal to the post-selected output. We can prepare this entangled state by introducing small rotation angles of $\pm\theta_0$ in the two paths of the interferometer, transforming the initial input state of $\ket{+}$ into
\begin{equation}
\ket{\psi} = \cos(\theta_0) \ket{+}\ket{V} + \sin(\theta_0) \ket{-}\ket{H}.
\end{equation}
With this small modification of the input state, $\braket{\psi|-}$ is orthogonal to $\braket{+|\psi}$ and the post-selected quantum Fisher information is given by 
\begin{equation}
    I_{ps} = \frac{1}{(\tan(\theta_0))^2}.
\end{equation}
Since the rotation angle $\theta_0$ can be arbitrarily small, the magnitude of the enhancement is only limited by experimental imperfections. 

To understand the mechanism by which the phase sensitivity gets transferred to photon polarization, it is useful to consider that polarization rotations can be represented by opposite phase shifts applied to the circular polarization components. The post-selected polarization state in the output port $\ket{-}$ can then be given as
\begin{align}
& \bra{-}\hat{U}(\phi)\ket{\psi} = 
 \nonumber \\
& \hspace{1cm} \frac{1}{\sqrt{2}}\left(\sin(\frac{\phi}{2}+\theta_0) \ket{R} - \sin(\frac{\phi}{2}-\theta_0) \ket{L}\right).
\label{eq:phistate}
\end{align} 
 The experimentally observed Fisher information is the classical Fisher information $I_{\mathrm{exp}}(\phi)$ of the phase dependent conditional probabilities $P(R|-)$ and $P(L|-)$. It can be determined from the gradient of the conditional probability $P(R|-)$,
\begin{equation}
\label{eq:IP}
    I_{\mathrm{exp}}(\phi) = \frac{\left(\displaystyle\frac{d}{d \phi} P(R|-)\right)^2}{P(R|-)\left(1-P(R|-)\right)}. 
\end{equation}
 In the absence of experimental imperfections, the phase dependence given by Eq.(\ref{eq:phistate})
has a classical Fisher information of
\begin{equation}
\label{eq:fullphase}
    I_{\mathrm{exp}}(\phi)=\left(\frac{\sin(2 \theta_0)}{1-\cos(2 \theta_0)\cos (\phi)}\right)^2.
\end{equation}
At $\phi=0$, the experimentally observed classical Fisher information is equal to the quantum Fisher information given in Eq. (\ref{eq:EFI}). As mentioned above, the post-selected quantum Fisher information depends on phase and reaches its maximum at $\phi=0$. This can be confirmed by examining the phase dependence of the experimentally observable Fisher information given by Eq.(\ref{eq:fullphase}). For $\phi \ll 1$, the Fisher information can be approximated by a Lorentzian,
\begin{equation}
\label{eq:Iapprox}
    I_{\mathrm{exp}}(\phi) \approx \frac{I_{ps}}{1+\phi^2/(2 \theta_0^2)}.
\end{equation}
At $\phi=\pm \theta_0$, the post-selected Fisher information is still equal to $2/3$ of the maximal value $I_{ps}$.

A direct comparison with multi-photon fringes is possible by taking the phase gradient at $P(R|-)=1/2$. For an $N$-photon fringe, the gradient is equal to $N/2$. We can define an effective photon number difference $A_{\mathrm{eff}}$ of
\begin{equation}
\label{eq:Aeff}
    A_{\mathrm{eff}} =  2 \left|\frac{d}{d\phi} P(R|-) \right| \Bigg|_{P(R|-)=1/2}.
\end{equation}
The enhanced phase sensitivity is then given by $I_{ps}=A_{\mathrm{eff}}^2$, the squared value of the effective photon number difference between the paths.

\begin{figure}[h]
    \centering
    \includegraphics[width=0.9\linewidth]{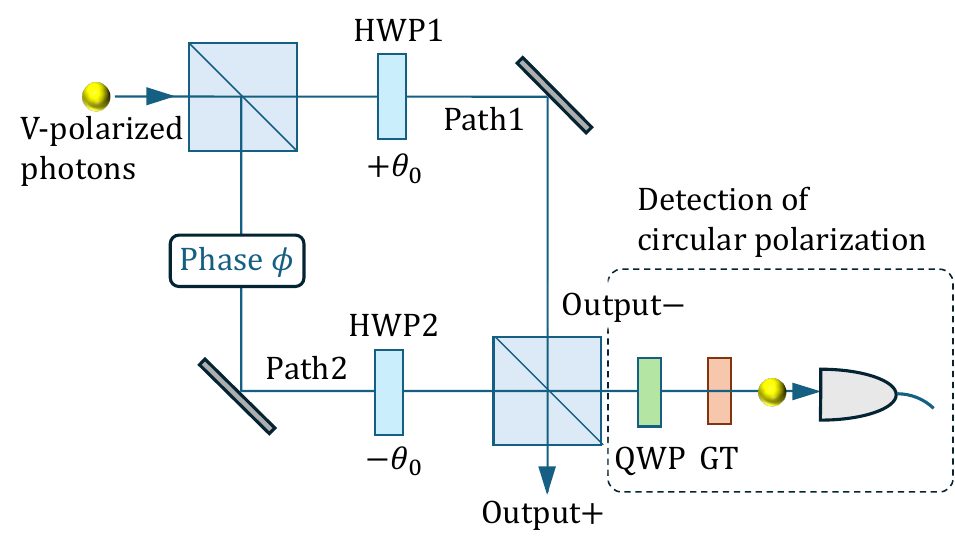}
    \caption{Schematic representation of the experimental setup for post-selection enhancement of single photon phase sensitivity in a conventional two-path interferometer. Single photons are initially polarized vertically(V-polarization) and injected into the $\ket{+}$ superposition of the paths. The initial polarization is locally rotated in each path by two half-wave plates (HWP1 and HWP2). HWP1 slightly rotates the polarization by $+\theta_0$, while HWP2 applies an equal rotation in the opposite direction $-\theta_0$. $\theta_0$ should be kept small to limit the associated loss of coherence. The small polarization rotations transfer phase sensitivity to the circular polarizations. When post-selecting the low probability output port $\ket{-}$, the phase gradient of the conditional probability $P(R|-)$ that the photon detected in $\ket{-}$ has right circular polarization is much higher than the value of 1/2 associated with the standard quantum limit of single photon interference. $P(R|-)$ was evaluated through separate counts of right and left polarized photons, which were accumulated for 50 s each for every value of $\phi$.
    }
\label{fig:setup}
\end{figure}

The transfer of phase sensitivity to the ancillary polarization degree of freedom via opposite polarization rotations in the two paths can be implemented in any existing two-path interferometer. Fig.~\ref{fig:setup} shows a schematic illustration of the experimental setup. In this illustration, the paths are separated in space to better indicate the operation of the optical elements. In the actual experiment, we used a Sagnac-type interferometric setup to obtain better stability, but the basic operations are the same as those shown in the figure. A semiconductor laser with a wavelength of 808.5 nm was first attenuated to approximately $2\times10^5$/s using neutral-density (ND) filters. The photons were polarized vertically (V state) using a polarizer and injected into the interferometer. A single beam splitter was used to implement interference between photons circulating right and photons circulating left in a Sagnac configuration. A slight spatial offset allowed us to apply a phase shift of $\phi$ in one path only by tilting a glass plate inserted in this path. We applied phase shifts over a range of $-\pi/8 \leq \phi \leq 9\pi/8$ in steps of $\pi/64$, corresponding to phase angles of $-22.5^{\circ} \leq \phi \leq 202.5^{\circ}$ in steps of $2.81^{\circ}$. 


The polarization rotations in the paths were implemented by placing two half-wave plates (HWP1 and HWP2) into the two paths. To obtain optimal results, the reduction of visibility by the polarization rotation should be larger than the losses of visibility due to experimental imperfections in the setup. Since the angle of polarization rotation $\theta_0$ is the essential experimental parameter controlling the magnitude of the expected effect, we took care to determine its precise value experimentally. We did so by blocking one path to observe the probability at which the rotation generated $H$-polarized output photons. The experimental result gave us an angle of $\theta_0=0.16848 \pm 0.00004$ or about $9.65^\circ$. The expected loss of visibility at this rotation angle is expected to be $2 \sin^2(\theta_0) = 0.05624$. The actual visibility observed in $\ket{-}$ was $V_-=0.9266 \pm 0.0005$, showing a slightly higher loss of visibility due to experimental imperfections. Under ideal conditions, the polarization rotation that we applied could result in an enhancement of Fisher information by a factor of $1/\tan^2(\theta_0)=34.56$, close to a six photon NOON state. However, this value will be reduced by the imperfections responsible for the additional reduction of visibility.

\begin{figure}[]
    \centering
    \includegraphics[width=1\linewidth]{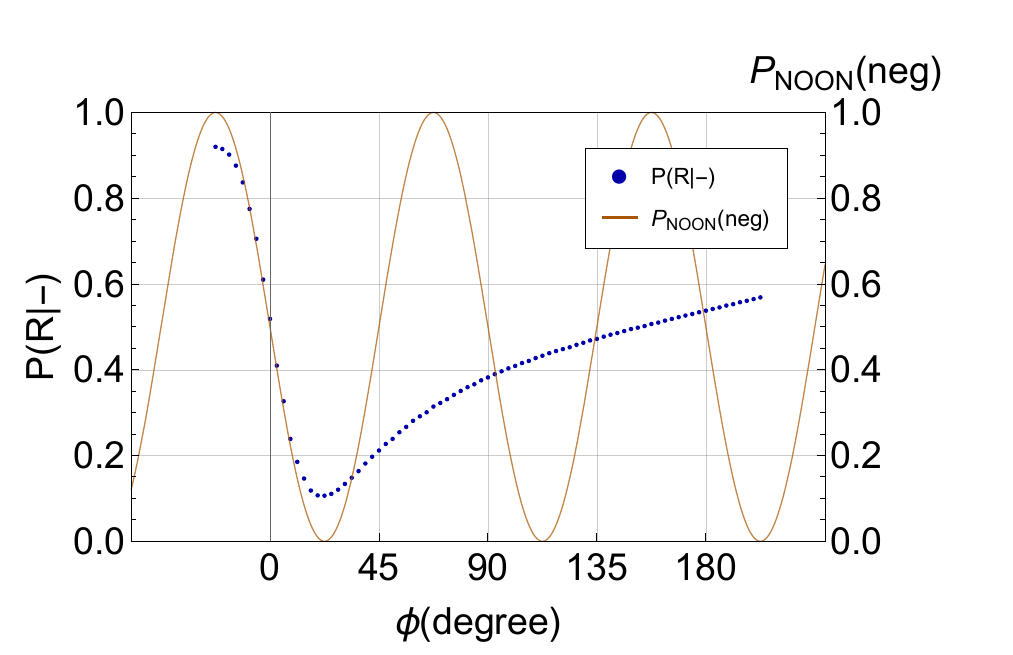}
    \caption{Conditional probability $P(R|-)$ of detecting the post-selected photons in the right circular polarization $R$ as a function of phase $\phi$. The solid blue circles represent the experimental results. Enhanced phase sensitivity is observed near $\phi=0$, where the post-selection probability $P(-)$ is minimal. The slope of $P(R|-)$ at $\phi\approx0$ is $-2.028\pm0.014$, which was obtained by averaging the values of three pairs, $\phi = (-\pi/64,\pi/64)$ and $(0,\pm\pi/32)$.  Taking into account the slight reduction in slope as the curve bends toward its extrema, the slope of $(0,\pm\pi/32)$ was corrected by a factor of $\pi/3$. The enhancement of phase sensitivity can be visualized by comparison with a four-photon fringe predicted for the parity measurement of an ideal NOON state. Here, $P_{\mathrm{NOON}}(\mbox{neg})$ indicates the probability of negative parity outcomes of the $N$-photon measurement. The phase sensitivity of the four photon NOON state is determined by the slope of $-2.0$ at $\phi=0$. It is almost indistinguishable from the slope of our post-selected experimental data. The statistical errors of each data point are below $2\times10^{-3}$ and smaller than the solid circles representing the data points.
    \label{fig:PR}}
\end{figure}

As explained in the theory section above, the enhanced phase sensitivity of photons detected in the low probability output port $\ket{-}$ is observed in the phase gradient of the conditional probability $P(R|-)$ of finding the photons in the right circular polarization $R$ at a phase shift close to zero, where the probability $P(-)$ of finding photons in the output $\ket{-}$ is minimal. Fig.~\ref{fig:PR} shows the conditional probability $P(R|-)$ of the post-selected photons as a function of the phase $\phi$. The slope at $\phi=0$ is much steeper than the maximal slope of $-1/2$ ($-1/2$ per $57.3^\circ$) observed without post-selection at a visibility of one. 

The phase sensitivity was evaluated based on the slope of $P(R|-)$ at $\phi=0$. To reduce the effects of errors, we determined the slope from the probability differences between points separated by two steps, or $\pi/32$. From the probability difference between $\phi=-\pi/64$ and $\phi=\pi/64$, we obtain a slope of $-2.043 \pm 0.023$. For comparison, we also obtained the probability differences between $\phi=-\pi/32$ and $\phi=0$, and between $\phi=0$ and $\phi=\pi/32$. The average slope between these points is expected to be lower than that around $\phi=0$, reflecting the decrease of the gradient as the probability approaches its extremal values. From the raw data, we get a slope of
$-1.903 \pm 0.025$ and $-1.957 \pm 0.020$, respectively. Comparison with an ideal interference fringe shows that these values should be a factor of $\pi/3$ lower than the gradient at $\phi=0$. Correction with this factor results in a slope of $-1.993 \pm 0.026$ and $-2.049 \pm 0.021$. The difference between the two results suggests that the value obtained at $\phi=0$ might have been a little bit too high. Otherwise, the three results are consistent with an average of $-2.028 \pm 0.014$.  

According to Eq.(\ref{eq:IP}) and (\ref{eq:Aeff}), a phase gradient of $-2.028$ at $P=1/2$ corresponds to a Fisher information of $I_{ps}=16.5$ and an effective photon number difference of $A_{\mathrm{eff}}=4.06$. The post-selected phase sensitivity of the photons detected in the low probability output port of our two-path interferometer is nearly as high as that of a four photon NOON state. Fig. ~\ref{fig:PR} shows the comparison of the experimental data with a four photon fringe, illustrating how close the post-selected single photon phase sensitivity gets to a phase sensitivity that would require at least four photons in a highly non-classical state without post-selection. 
The two curves in Fig.\ref{fig:PR} overlap in a phase interval between $-\theta_0
=-9.65\degree$ and $+\theta_0=+9.65\degree$, which corresponds to the interval in which the enhanced Fisher information is expected to be greater than $2/3 I_{ps}$ according to Eq.(\ref{eq:Iapprox}). For an ideal NOON state fringe, the Fisher information is the same at all phases, so that the visual comparison in Fig. \ref{fig:PR} is a good indicator of the range of phase values that achieve near optimal enhancements. The reduction of Fisher information described by Eq.(\ref{eq:Iapprox}) can be observed in the slightly higher curvature that distinguishes the experimental data from the ideal NOON state fringe near $\phi=0$. It is therefore reasonable to focus on the slope of $P(R|-)$ at $\phi=0$ in the evaluation of post-selected phase sensitivity. We would like to emphasize that we were able to confirm the phase sensitivity of our setup directly from the experimental count rates, without any corrections for losses or imperfections. As noted in \cite{Slu17}, most of the results on NOON state fringes achieve much less than their optimal sensitivity without such corrections. It is one of the advantages of post-selected metrology that the Fisher information is based entirely on actually detected events, requiring no assumptions about detection efficiencies and other photon losses in the setup.

The effective value $A_{\mathrm{eff}}$ of the photon number difference between the paths expressed by the operator $\hat{A}$ is directly determined by the slope of the post-selected probabilities at $P(R|-)=1/2$. In our experiment, the phase sensitivity corresponds to a photon number difference of $A_{\mathrm{eff}}=4.06$. It is possible to interpret this anomalous value of $A_{\mathrm{eff}}$ as a conditional uncertainty of $\hat{A}$, corresponding to the formulation of uncertainty by Ozawa and Hall \cite{Hof11,Oza03,Hal04,Ji26}. In our previous work, we observed this conditional uncertainty directly and identified the enhanced value of $\hat{A}$ with a super-localization effect conditioned by the observation of a photon in a low probability output port \cite{Fuk26}. The method introduced here was inspired by this result and the realization that the anomalous conditional uncertainty reported in \cite{Fuk26} corresponds directly to the negative quasiprobability identified as the source of enhanced phase sensitivity in \cite{Arv20}. The present work shows that the relation between phase sensitivity and the uncertainty of photon number difference between the paths of an interferometer extends also to post-selected phase estimation, where anomalous conditional uncertainties can enhance phase sensitivity beyond the Heisenberg limit defined by the maximal input state uncertainty without post-selection \cite{Hof11,Ji26}. Conditional uncertainties are not limited by the eigenvalues of their operators, allowing us to transcend the uncertainty limits that apply to quantum states by applying an appropriate post-selection procedure. 

In summary, we have introduced a method to activate the phase sensitivity concentrated in the low probability output port of a two-path interferometer by transferring the phase sensitivity to the circular polarization of the photons before post-selecting the photons exiting the dark port of the interferometer. This method makes it possible to post-select only a single output port of any two path interferometer without losing the phase sensitivity concentrated in that port. Our results show that the fundamental physics of single photon interference has practical applications in conventional two-path interferometers, requiring only minimal modifications to access the full potential of its non-classical statistics. 

This work was supported by ERATO, Japan Science and Technology Agency (JPMJER2402), JST SPRING, Grant No. JPMJSP2132 and JSPS KAKENHI Grant Number JP26K08231

The data used to generate figure \ref{fig:PR} are available from Ref. \cite{data}.

\nocite{*}

\end{document}